\documentclass[aps,prl,twocolumn,showpacs,floatfix,superscriptaddress]{revtex4}
\usepackage{amsmath,amsbsy,amssymb,latexsym}
\usepackage{graphicx,float}
\usepackage{color}

\def\la{\langle}
\def\ra{\rangle}
\def\ve{\vert}

\begin{document}
\title{Statistical mechanics of nucleosome ordering by chromatin structure-induced two-body interactions}

\author{R\u{a}zvan V. Chereji}
\affiliation{Department of Physics and Astronomy, Rutgers University, Piscataway, NJ 08854-8019}

\author{Denis Tolkunov}
\affiliation{Department of Physics and Astronomy, Rutgers University, Piscataway, NJ 08854-8019}
\affiliation{BioMaPS Institute for Quantitative Biology, Rutgers University, Piscataway, NJ 08854-8019}

\author{George Locke}
\affiliation{Department of Physics and Astronomy, Rutgers University, Piscataway, NJ 08854-8019}

\author{Alexandre V. Morozov}
\thanks{Corresponding author: morozov@physics.rutgers.edu}
\affiliation{Department of Physics and Astronomy, Rutgers University, Piscataway, NJ 08854-8019}
\affiliation{BioMaPS Institute for Quantitative Biology, Rutgers University, Piscataway, NJ 08854-8019}

\date{\today}

\begin{abstract}
One-dimensional arrays of nucleosomes (DNA-bound histone octamers separated by stretches of linker DNA)
fold into higher-order chromatin structures which ultimately make up eukaryotic chromosomes.
Chromatin structure formation leads to $10-11$ base pair (bp) discretization of linker lengths caused by the smaller free energy
cost of packaging nucleosomes into regular chromatin fibers if their rotational setting (defined by the DNA helical twist) is conserved.
We describe nucleosome positions along the fiber using a thermodynamic model of finite-size particles
with both intrinsic histone-DNA interactions and an effective two-body potential. 
We infer one- and two-body energies directly from high-throughput maps of nucleosome positions.
We show that chromatin structure explains \textit{in~vitro} and \textit{in~vivo} nucleosome ordering
in transcribed regions, and plays a leading role in establishing well-known $10-11$~bp genome-wide
periodicity of nucleosome positions.
\end{abstract}

\maketitle

In living cells, eukaryotic DNA is found in a compact, multi-scale chromatin state \cite{Felsenfeld:2003}.
The fundamental unit of chromatin is a nucleosome: 
$147$~bp of DNA wrapped around a histone octamer \cite{Richmond:2003}.
In addition to its primary function of DNA compaction, chromatin modulates DNA accessibility
to transcription factors and other molecular machines in response to external signals, 
exerting a profound influence on numerous DNA-mediated biological processes such as gene transcription, DNA repair, and replication~\cite{Li:2007}.

Equilibrium thermodynamic models that account for intrinsic histone-DNA sequence preferences and nearest-neighbor steric exclusion have been used to
predict nucleosome positions and formation energies \cite{Morozov:2009,Kaplan:2009,Locke:2010}.
However, structural regularity of the chromatin fiber imposes additional constraints,
leading to discretization of linker lengths between neighboring nucleosomes with the $10-11$~bp periodicity of the DNA double helix \cite{Strauss:1983,Widom:1992}.
The discretization is required to avoid steric clashes caused by the nucleosome rotating around the linker DNA axis as the linker is extended \cite{Ulanovsky:1986},
and more generally to minimize the free energy costs associated with maintaining a regular pattern of protein-protein and protein-DNA contacts
in the chromatin fiber \cite{Widom:1992}. Indeed, adding a short DNA segment to the linker will result in a rotation of the nucleosome
with respect to the rest of the fiber, disrupting its periodic structure.
This additional twist has to be compensated unless the segment is $10-11$~bp in length, bringing the nucleosome into an equivalent rotational position.

Large-scale maps of \textit{in~vivo} and \textit{in~vitro} nucleosome positions in yeast reveal nucleosome-depleted regions (NDRs)
in the vicinity of transcription start and termination sites (TSS and TTS) \cite{Mavrich:2008, Kaplan:2009, Zhang:2009}.
In these experiments, chromatin is digested with micrococcal nuclease to obtain mononucleosome core 
particles, and the mononucleosomal DNA is purified and either sequenced or hybridized to microarrays \cite{Tolkunov:2010}.
5' NDRs play a key role in gene regulation \cite{Mavrich:2008}.
NDRs are also observed \textit{in~vitro}, where they are defined by poly(dA:dT) tracts and other nucleosome-disfavoring sequences.
Surprisingly, there are no oscillations in nucleosome occupancy around \textit{in vitro} NDRs and,
on average, just a $\sim 25\%$ depletion of the occupancy over 5' NDRs compared with the genome-wide mean \cite{Kaplan:2009, Zhang:2009}
(bp occupancy is defined as its probability to be nucleosome-covered).
This is true even if genomic DNA from \textit{S.cerevisiae} is combined with purified histones in a 1:1 mass ratio,
yielding a maximum nucleosome occupancy of $0.82$ which is close to the \textit{in~vivo} value \cite{Zhang:2009}. 
This behavior is in sharp contrast with \textit{in~vivo} chromatin in which the action of transcription factors,
chromatin remodeling enzymes and components of transcriptional machinery results in well-positioned genic nucleosomes
and highly pronounced 5' NDRs ($\sim 70\%$ depletion on average with respect to the mean) \cite{Mavrich:2008,Kaplan:2009}.
Because occupancy oscillations are a generic feature of  one-dimensional liquids of finite-size particles in the vicinity of potential
barriers and wells \cite{Kornberg:1988}, the absence of such oscillations \textit{in~vitro} and shallow NDRs strongly suggest that
sequence-specific histone-DNA interaction energies are on average comparable to~$k_B T$.
Consequently, nucleosome-positioning and disfavoring sequences are expected to play a minor role in establishing \textit{in~vivo} localization
of genic nucleosomes.

Here we focus on how nucleosome positions are affected by effective two-body interactions imposed on neighboring particles by regular chromatin structure. 
We map a three-dimensional chromatin fiber onto a system of non-overlapping particles of length $a=147$~bp with both
histone-DNA and short-range nearest-neighbor interactions. The particles are confined to a one-dimensional lattice of length $L$.
We develop a theory in which the interaction 
(that reflects linker discretization) is deduced exactly from the two-particle distribution,
even in the presence of $10-11$~bp periodic one-body energies related to the rotational positioning of the nucleosome \cite{Mavrich:2008, Zhang:2009}.

\begin{figure}[t]
  \includegraphics[width=3.4in]{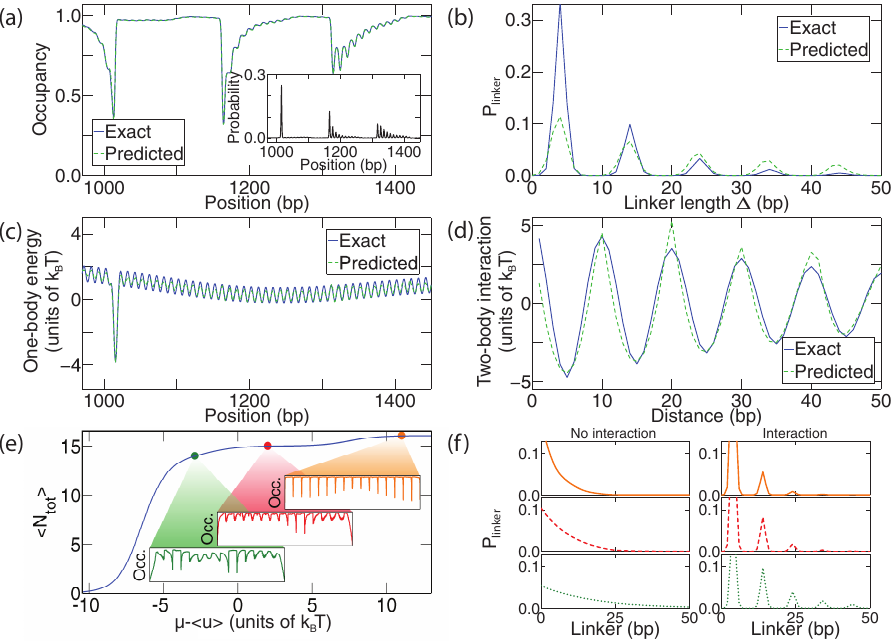}
  \caption{\label{Fig:model10bp} (Color online) A model with $10$~bp oscillations in both one-body and two-body energies.
The two-body interaction is $\Phi(x) = A \cos \left( \frac{2 \pi}{10} x \right) e^{-x/b}$,
where $A=5~k_B T$ and $b = 50 \text{ bp}$. For the one-body potential,
$10$~bp oscillations with the $0.5~k_B T$ amplitude were superimposed onto a smooth energy profile with
two  $-5~k_B T$ potential wells separated by $1000$~bp. DNA length of $2416$~bp was chosen to be able to position $16$ nucleosomes with $151$~bp repeat length.
The occupancy profile (a), the linker length distribution (b), the one-body energy (c),
and the two-body interaction (d): exact (solid blue line) and predicted (dashed green line).
$\mu - \la u \ra = -1~k_B T$ in (a)-(d).
Inset of (a): the probability of starting a nucleosome at a given bp.
(e) Average number of nucleosomes $\la N_{tot} \ra$ vs. $\mu - \la u \ra$.
Insets: Occupancy profiles corresponding to three different chemical potentials.
(f) Linker length distributions for three values of $\la N_{tot} \ra$ shown as points in (e), with and without two-body interactions.}
\end{figure}

Let $u(k)$ be the external potential energy of a particle that occupies positions $k$ through $k + a - 1$ on the DNA,
and let $\Phi(k,l)$ be the two-body interaction between a pair of nearest-neighbor particles with starting positions $k$ and $l$, respectively.
Here $u(k)$ describes intrinsic histone-DNA interactions, whereas $\Phi(k,l)$ accounts for the effects of chromatin structure.
The grand-canonical partition function is given by
\begin{equation}
	Z = 1 + \sum_{N=1}^{N_{max}} \la J\ve (z w)^{N-1} z \ve J\ra = 1 + \la J \ve (I-z w)^{-1} z \ve J \ra, \label{Eq:Z}
\end{equation}
where $N_{max}$ is the maximum number of particles that can be positioned on $L$~bp, $I$ is the identity matrix, $\ve j \ra$ is a unit vector of dimension $L-a+1$ with $1$ at position $j$, and 
$\ve J \ra = \sum_{j=1}^{L-a+1} \ve j \ra$. In matrix notation, $\la k \ve z \ve l\ra = e^{\beta [\mu-u(k)]} \delta_{k,l}$ and $\la k \ve w \ve l\ra = e^{-\beta \Phi(k,l)} \Theta(l - k)$, where $\mu$ is the chemical potential, 
$\delta_{k,l}$ is the Kronecker delta, $\beta$ is the inverse temperature, and $\Theta$ is the Heaviside step function.

The one-particle and nearest-neighbor pair distribution functions are:
\begin{align}
	n(i) &= \frac{1}{Z} \la J \ve (I - z w)^{-1} \ve i \ra \la i \ve z \ve i \ra \la i \ve (I - w z)^{-1} \ve J \ra, \label{Eq:n}\\
	\overline{n}_2(i,j) &= \frac{1}{Z} \la J \ve (I - z w)^{-1} \ve i \ra \la i \ve z w z \ve j \ra \la j \ve (I- w z)^{-1} \ve J \ra. \label{Eq:n2}
\end{align}

Note that for $0<j-i<2a$, $\overline{n}_2(i,j)=n_2(i,j)$, where $n_2$ is the ordinary two-particle distribution function.
Defining two matrices, $\la i \ve N \ve j \ra = n(i) \delta_{i,j}$ and $\la i \ve N_2 \ve j \ra = \overline{n}_2(i,j)$, we rewrite the partition function as
\begin{equation}
 Z = \frac{1}{1 - \la J \ve (I- N_2 N^{-1}) N \ve J \ra}. \label{Eq:Z2}
\end{equation}

By inverting Eqs.~\eqref{Eq:n} and \eqref{Eq:n2} we obtain the exact expressions for one- and two-body energies \cite{Percus:1976, Percus:1989}:
\begin{widetext}
\begin{align}
-\beta \left[u(k)-\mu\right]&= \ln \left(\frac{\la J \ve I - N_2 N^{-1} \ve k \ra \la k \ve N \ve k \ra \la k \ve I- N^{-1} N_2 \ve J \ra}{1 - \la J \ve I- N_2 N^{-1} N \ve J \ra}\right), \label{Eq:u}\\
-\beta \Phi(k,l)&= \ln \left(\frac{\la k \ve N^{-1} N_2 N^{-1} \ve l \ra \left[ 1 - \la J \ve I- N_2 N^{-1} N \ve J \ra \right]}
	{\la k \ve I- N^{-1} N_2 \ve J \ra \la J \ve I - N_2 N^{-1} \ve l \ra}\right). \label{Eq:Phi}
\end{align}
\end{widetext}
Note that if the two-body interactions are neglected, Eq.~\eqref{Eq:u} reduces to~\cite{Locke:2010}
\begin{equation}
e^{-\beta \left[u(i)-\mu\right]} = \frac{n(i)}{1 - O(i) + n(i)} \prod_{j=i}^{i+a-1} \frac{1 - O(j) + n(j)}{1 - O(j)}, \label{Eq:u0}
\end{equation}
where $O(i)$ is the nucleosome occupancy of bp $i$ [$O(i) = \sum_{j = i-a+1}^{i} n(j)$].

If one-body energies $u$ and two-body interactions $\Phi$ are known, Eqs.~\eqref{Eq:n} and \eqref{Eq:n2} allow us to construct
particle distributions $n$ and $\overline{n}_2$ exactly.
Conversely, we can use Eqs.~\eqref{Eq:u} and \eqref{Eq:Phi} to find $u$ and $\Phi$ from one- and two-particle distributions.
However, the two-particle distribution is not directly measured in current high-throughput experiments, in which
chromatin from many cells is mixed together before mononucleosomes are isolated and sequenced.
In other words, it is not known which particular genome a given nucleosome comes from.
This is irrelevant for $n$ but may present a problem for $\overline{n}_2$, which requires two-nucleosome configurations.
Nonetheless, we can build a model for $\overline{n}_2$ which allows us to approximate the two-body interaction.

Let $g(i,j)$ be the pair distribution $n_2(i,j)/[n(i) n(j)]$. 
Without one-body energies, the system is homogeneous and $g$ is a function of only the relative distance between the nucleosomes:
$g(i,j)\equiv g(j-i)$. In this case Eq.~\eqref{Eq:Phi} reduces to
\begin{equation}
-\beta \Phi (i,j) = \ln \left[g(j-i)\right] + \alpha (j-i) + \ln C
\end{equation}
for arbitrary interactions $\Phi$ \cite{Gursey:1950}. The constants $C$ and $\alpha$ can be determined
from the asymptotic condition $\lim_{(j-i) \to \infty} \Phi(i,j)= 0$.
However, position-dependent one-body energies break translational invariance of the pair distribution $g$.
Assuming that $\Phi$ is translationally invariant, 
we introduce $P_{\text{linker}} (\Delta) = \la g(i,i+\Delta+a) \ra_i$ and approximate $\Phi$ as
\begin{equation} \label{Eq:Phi:est}
-\beta \Phi (i,j) \approx \ln \left[P_{\text{linker}}\boldsymbol{(}j-(i+a)\boldsymbol{)}\right] + \alpha (j-i) + \ln C.
\end{equation}
This step is reminiscent of replacing the ensemble average with the time average in statistical mechanics.
Our numerical tests show this to be an excellent approximation, even if one- and two-body energies are comparable in magnitude, making
the system strongly inhomogeneous.

Experimental nucleosome positioning data sets consist of the histogram of the number of nucleosomes starting at each genomic bp $i$.
We preprocess the data by removing all counts of height $1$ from the histogram, and smoothing
the remaining profile with a $\sigma=2$ Gaussian kernel.
Next, we compute $n(i)$ by rescaling the smoothed profile so that the maximum occupancy for each chromosome is $1$.
Finally, we identify all local maxima on the $n$ profile and assume that they mark prevalent nucleosome positions.
For each maximum at bp $i$ we find subsequent maxima at positions $i+146<j_1<j_2<j_3<\dots$ in the $50$ bp window.
To each pair of maxima $(i,j_1),~(i,j_2), \dots$ we assign the probability that they represent neighboring nucleosomes: $n(i) n(j_1), ~n(i) [1-n(j_1)] n(j_2), \dots$
By summing over all initial positions $i$ and normalizing, we obtain the linker length probability which
gives us an empirical estimate of $P_{\text{linker}}$.

Fig.~\ref{Fig:model10bp} demonstrates our procedure in a model system, with preprocessing and rescaling steps skipped
since the simulated $n$ profile is noise-free and already properly normalized. Specifically, we use local maxima in the nucleosome starting probability profile
[inset of Fig.~\ref{Fig:model10bp}(a)] to obtain $P_{\text{linker}}$ [Fig.~\ref{Fig:model10bp}(b)].
Fig.~\ref{Fig:model10bp}(d) shows that the two-body interaction can be reconstructed using Eq.~\eqref{Eq:Phi:est},
even in the presence of one-body energies with the same periodicity.
The reconstruction is facilitated by the presence of potential wells or barriers in the one-body energy profile that are strong enough
to create non-uniform density of nearby nucleosomes.
To find the one-body energies, we substitute predicted $\Phi$ into Eq.~\eqref{Eq:n},
which we solve numerically for $z$ [Fig.~\ref{Fig:model10bp}(c)]. Nucleosome occupancies inferred from predicted $u$ and $\Phi$
are virtually identical to the exact profile~[Fig.~\ref{Fig:model10bp}(a)].

As the chemical potential is increased, nucleosomes undergo a transition in which their average number
goes up in a step-like fashion [Fig.~\ref{Fig:model10bp}(e)] \cite{Schwab:2008}.
In contrast to the $\Phi = 0$ case in which linkers are distributed exponentially, two-body interactions lead to the pronounced discretization
of linker lengths [Fig.~\ref{Fig:model10bp}(f)]. The first minimum of $\Phi$ becomes more dominant as the number of nucleosomes increases,
leading to a well-positioned array with $4$~bp-long linkers.

\begin{figure}[t]
  \includegraphics[width=3.4in]{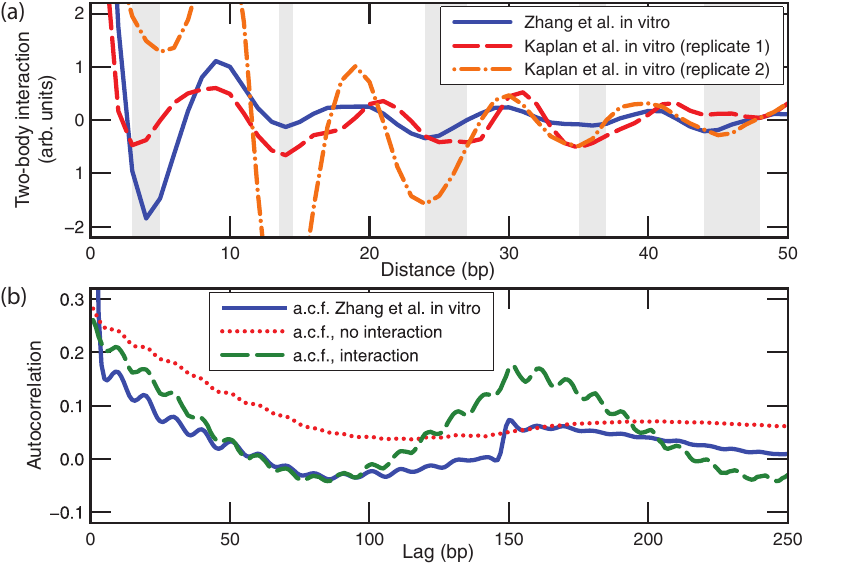}
  \caption{\label{Fig:datasets} (Color online) (a) Two-body interaction $\Phi$ inferred from \textit{in~vitro} maps of
nucleosome positions \cite{Kaplan:2009,Zhang:2009}.
Grey bars indicate consensus positions of the minima.
(b) Autocorrelation of nucleosome starting positions in one of the \textit{in~vitro} data sets~\cite{Zhang:2009},
and of starting positions predicted using sequence-specific one-body energies from the ``spatially resolved'' model~\cite{Locke:2010},
with and without $\Phi$. The two-body potential is from Fig.~\ref{Fig:model10bp},
consistent with the minima of $\Phi$ observed in (a). The one-body energies have $\sigma = 0.23~k_B T$.
To account for the limited size of the \textit{in~vitro} data set, model output was degraded by randomly removing $1\%$ of predicted
nucleosome probabilities.
}
\end{figure}

We now use Eq.~\eqref{Eq:Phi:est} to predict nearest-neighbor interactions from genome-wide nucleosome maps
[Fig.~\ref{Fig:datasets}(a)]. We find that despite significant experiment-to-experiment variations,
all two-body potentials have minima within $1-2$~bp of $5+10m \text{ bp},~m=0,1,\dots$ \cite{Wang:2008}.
Surprisingly, there are substantial differences between two Kaplan \textit{et al.} \cite{Kaplan:2009} \textit{in~vitro} replicates,
with one replicate exhibiting higher values of $\Phi$ due to the pronounced depletion of nucleosomes separated by $<10 \text{ bp}$.
Apparently, chromatin structure can undergo subtle uncontrolled changes from experiment to experiment.

\begin{figure}[t]
  \includegraphics[width=3.4in]{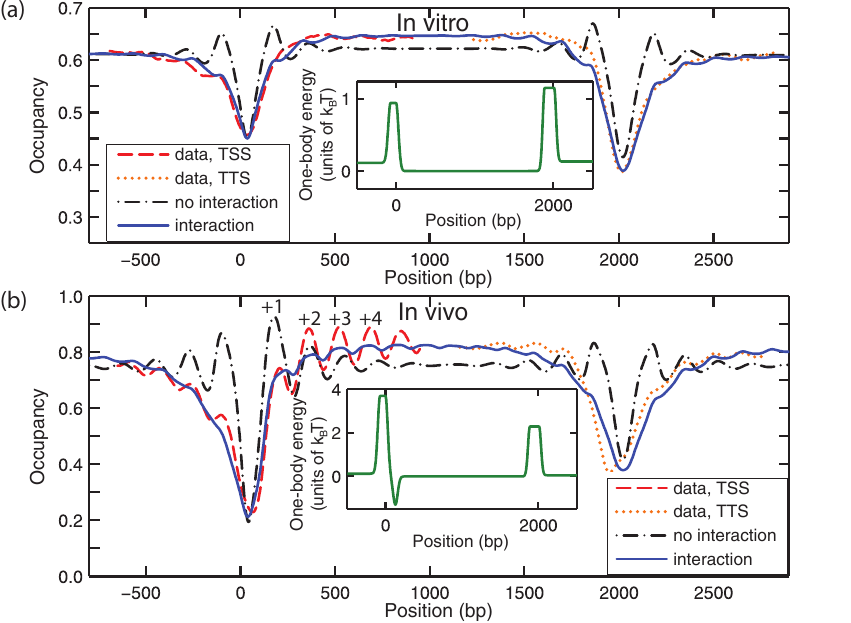}
  \caption{\label{Fig:models} (Color online) A minimal model of nucleosome ordering in genic regions.
(a) Dashed red and dotted orange lines: average nucleosome occupancy \textit{in~vitro} around TSS and TTS \cite{Zhang:2009}.
Solid blue and dash-dot black lines: model predictions with and without $\Phi$
from Fig.~\ref{Fig:model10bp}. Both models have the average occupancy of $0.60$
(less than the maximum possible occupancy of $0.82$ because some histone octamers are not DNA-bound).
Inset: one-body energy landscape with barrier heights, widths and shapes adjusted to reproduce observed NDRs.
(b) Same as (a), for \textit{in~vivo} nucleosomes (YPD medium) \cite{Zawadzki:2009}.
$\Phi$ is from Fig.~\ref{Fig:model10bp} with $A=7~k_B T$.
The log-intensities from the microarray were exponentiated and normalized separately for each gene,
yielding the average occupancy of $0.70$ which was also used in the models.}
\end{figure}

Two-body interactions are reflected in the autocorrelation of nucleosome starting positions [Fig.~\ref{Fig:datasets}(b)].
The oscillations in the autocorrelation function are suppressed when nucleosome positions are predicted using
a sequence-specific model which neglects two-body interactions~\cite{Locke:2010}.
This ``spatially resolved'' model assigns mono- and dinucleotide energies independently at each position
within the nucleosomal site and is thus capable of capturing the $10-11$~bp periodicity of one-body interactions.
We find that the autocorrelation function is much closer to experiment
if the two-body potential is included into the model [Fig.~\ref{Fig:datasets}(b)].

Two-body interactions are also essential for reconstructing nucleosome occupancy profiles over
transcribed regions [Fig.~\ref{Fig:models}].
Sequence-specific energy barriers over NDRs must be low \textit{in~vitro} to account for the lack of occupancy oscillations
induced by steric exclusion at 1:1 DNA:histone mass ratio \cite{Zhang:2009}. Even with the low barriers shown in Fig.~\ref{Fig:models}(a),
the interaction-free model yields an oscillatory profile which is not observed in the data. The oscillations are suppressed by the two-body potential,
and the resulting profile increases towards the center of the gene, in contrast with the pure steric exclusion scenario in which nucleosomes adjacent
to the barriers are always the most localized \cite{Kornberg:1988}. This behavior 
is also observed \textit{in~vivo} where the $+2$ nucleosome is higher than the $+1$ nucleosome [Fig.~\ref{Fig:models}(b)].
The \textit{in~vivo} barriers are more pronounced
to account for additional nucleosome depletion in the NDRs due to effects other than intrinsic histone-DNA interactions.
Finally, in agreement with a previous hypothesis \cite{Zhang:2009}, a potential well is added to localize the $+1$ nucleosome \textit{in~vivo}.
The well makes the TSS profile asymmetric with respect to the center of the NDR [compare to the more symmetric TTS profile in Fig.~\ref{Fig:models}(b)].

In summary, our study is the first to show that short-range two-body interactions induced by chromatin fiber formation
play a major role in genome-wide nucleosome ordering. We demonstrate that
large-scale mononucleosome maps contain evidence of the two-body potential.
This potential is more important than intrinsic histone-DNA interactions
for predicting $10-11$~bp periodicity in genome-wide nucleosome positions, and for understanding
nucleosome occupancy in transcribed regions.
Clearly, two-body interactions should be
an integral part of genome-wide models of nucleosome occupancy.
Our study also underscores the need for future experiments focused on multi-nucleosome distributions,
which can be analyzed using our exact theory [Eqs.~\eqref{Eq:u} and \eqref{Eq:Phi}].

This research was supported by National Institutes of Health (HG 004708) and by an Alfred P. Sloan Research Fellowship to AVM.

\end{document}